\def\aa{A\&A}            %Astronomy & Astrophysics%
\def\anj{AJ}             %Astronomical Journal%
\def\apj{ApJ}            %Astrophysical Journal%
\def\baas{BAAS}          %Bulletin of the American A.S.%
\def\mn{MNRAS}           %Monthly Notices of the Royal...%
\documentclass[]{aanew}

\begin{document}

\title{Variability and polarization in the inner jet of 3C\,395}

\author{L. Lara\inst{1} \and 
A. Alberdi\inst{1} \and
J.M. Marcaide\inst{2}}

\offprints{L. Lara, \email{lucas@iaa.csic.es}}

\institute{Instituto de Astrof\'{\i}sica de Andaluc\'{\i}a (CSIC),
Apdo. 3004, 18080 Granada (Spain)
\and 
Departamento de Astronom\'{\i}a, Universitat de Val\`encia, 46100 Burjassot  
(Spain)} 
\date{Received / Accepted}

\authorrunning{Lara et al.}
\titlerunning{Variability and polarization in the inner jet of 3C\,395}

\abstract{
We present new results on the parsec-scale jet of the quasar 3C\,395,
derived from VLBI polarization sensitive observations made in 1995.91
and 1998.50 at 8.4, 15.4 and 22.2 GHz. The observations show a complex
one-sided jet extending up to 20 mas, with a projected magnetic field
essentially aligned with the radio jet. The emission is strongly
dominated, in total intensity and polarization, by the core and the
inner jet region (of $\sim 3$ mas length). We have studied the details
of this dominant region finding clear structural variations during
this $\sim 2.5$ years period, in contrast with the apparent quietness
of the jet structure inferred from lower resolution VLBI
observations. We observe the ejection of a new component from the core
and variations in the degree of polarization of the inner jet
components. We estimate a high Faraday Rotation Measure close to the
core, with a strong decrease along the inner jet.
}

\maketitle

\section{Introduction}

The quasar \object{3C\,395} (1901+319; z=0.635) presents at milliarcsecond
scales a radio structure which basically consists of two components
(hereafter A and B), which are stationary with respect to each other
and have a separation of $\sim$16~mas, at a position angle of
118$^{\circ}$.  A third weak and extended component (C) appears
located between those two in maps made at frequencies lower than 15
GHz (Lara et al. \cite{lara1}, \cite{lara2}). A flux density
monitoring at 5.0, 8.0 and 14.5 GHz made by the University of Michigan
Radio Astronomy Observatory (UMRAO) shows that 3C\,395 exhibits strong
variability which presumably is the result of activity within
component A.  In fact, high resolution space Very Long Baseline
Interferometry (VLBI) observations resolve component A into a core-jet 
structure with
bends. The existence of this complex structure might clarify the apparent 
contradiction between the lack of observed structural changes
from lower resolution VLBI observations and the flux density
variability (Lara et al. \cite{lara3}). The stationary component B has 
been interpreted as the result of a local
bend in the jet towards the observer, while component C seems more
related to the emission of a complex underlying jet flow (see Lara et al.
\cite{lara1}, \cite{lara2}, \cite{lara3}).  Polarimetric observations of 
3C\,395 have recently been reported by Taylor (\cite{taylor1}), who
presents, in the framework of a study of Faraday Rotation Measure (RM)
in the inner jets of a sample of quasars, polarization sensitive
VLBA\footnote{Very Long Baseline Array, operated by the National Radio
Astronomy Observatory (NRAO).} observations at frequencies between 4.6
and 15.2 GHz. Taylor finds a strong gradient in the RM of component A,
with the highest value ($\sim +1200$ rad\,m$^{-2}$) at the western end
of this component.
  
It is now clear that the understanding of the activity and evolution
of 3C\,395 requires the study of the structure of component A at
sub-milliarcsecond resolution.  Accordingly, we present in this paper
polarization sensitive observations of the quasar 3C\,395 made with
the VLBA at 8.4, 15.4 and 22.2 GHz, with angular resolutions reaching
0.32 mas ($\simeq$1.24 pc\footnote{We assume H$_0 = 100$
Km\,s$^{-1}$\,Mpc$^{-1}$ and q$_0 = 0.5$ throughout.}) at 22.2 GHz.

\section{Observations and data reduction}

We carried out VLBA observations of the quasar 3C\,395 switching between 
frequencies at 8.4 and 15.4 GHz in November 27th 1995 (epoch 1995.91)
and at 15.4 and 22.2 GHz in June 30th 1998 (epoch 1998.50). Left and
right circular polarizations were recorded during both observing
runs. The synthesized bandwidths per circular polarization were 16 MHz
and 32 MHz in 1995 and 1998, respectively.

The correlation of the data was done {\em in absentia} by the staff of
the VLBA correlator in Socorro (NM, USA). We used the NRAO
AIPS\footnote{Astronomical Image Processing System, developed and
maintained by the NRAO.} package to correct for instrumental phase and
delay offsets between the separate baseband converters in each
antenna, and to determine antenna-based fringe corrections. The
visibility amplitudes were calibrated using the system temperatures
and gain information provided for each telescope. We estimate
amplitude calibration errors to be smaller than 5\% at all
observing frequencies.

In both epochs, the determination of the feed responses to the
polarized signal at each antenna was done using the feed solution
algorithm of Lepp\"anen et al. (\cite{leppanen}), which calculates the
so called D-terms (that is, the terms describing the ``leakage'' of
the orthogonal polarization into each feed). Independent determinations 
of the instrumental
polarization were done for each observed source,
including 3C\,395, obtaining a good agreement in the results at all frequencies and epochs.  This
allowed us to apply the corrections derived from a given source in
the polarization mapping of that source.
During the calibration, we assumed that the circularly
polarized emission from all the sources was negligible, as suggested by 
the data.
In 1995, the number of known sources suitable for
polarization calibration of VLBI observations at frequencies higher
than 15 GHz was very small. We took snapshots of the sources
\object{1656+053}, \object{2145+067}, \object{3C\,84} and \object{OQ\,208}, 
expecting to find a good
calibrator of the absolute orientation of the electric field vector
(electric vector position angle - EVPA) among them. 
The final EVPA was derived from the direct comparison of Very
Large Array (VLA; close-in-time data were requested from the VLA
public archive) and our VLBA polarization images of 1656+053, which
presents a compact structure at the angular resolutions provided by
both instruments. From the VLA data we determined an EVPA = $-15^{\circ}$  
for this source. This orientation was consistent with that derived
from the D-terms determined by Lepp\"anen from a close-in-time
VLBA experiment at 8.4 GHz (private communication). 
In 1998 the number of known 
polarization calibrators had increased considerably.  We
observed \object{3C\,279}, \object{1611+343} and \object{0420-014}, all with 
known polarization
properties, and we could obtain a consistent calibration for all of them.
The absolute EVPA was derived from 0420-014, comparing our VLBA 
observations with the UMRAO database, and from the outer component in 
3C\,279 (Taylor \& Myers \cite{taylor2}).  
The EVPA were determined with an error that we estimate to be within
$10^{\circ}$ at all epochs and frequencies.
 
Data imaging in total intensity was performed with the Difmap package
(Shepherd et al. \cite{shepherd}).  Maps of the Stokes parameters Q
and U were made and combined in AIPS to finally obtain maps of the linearly
polarized emission of 3C\,395.

\begin{figure*}
\vspace{23cm}
\includegraphics{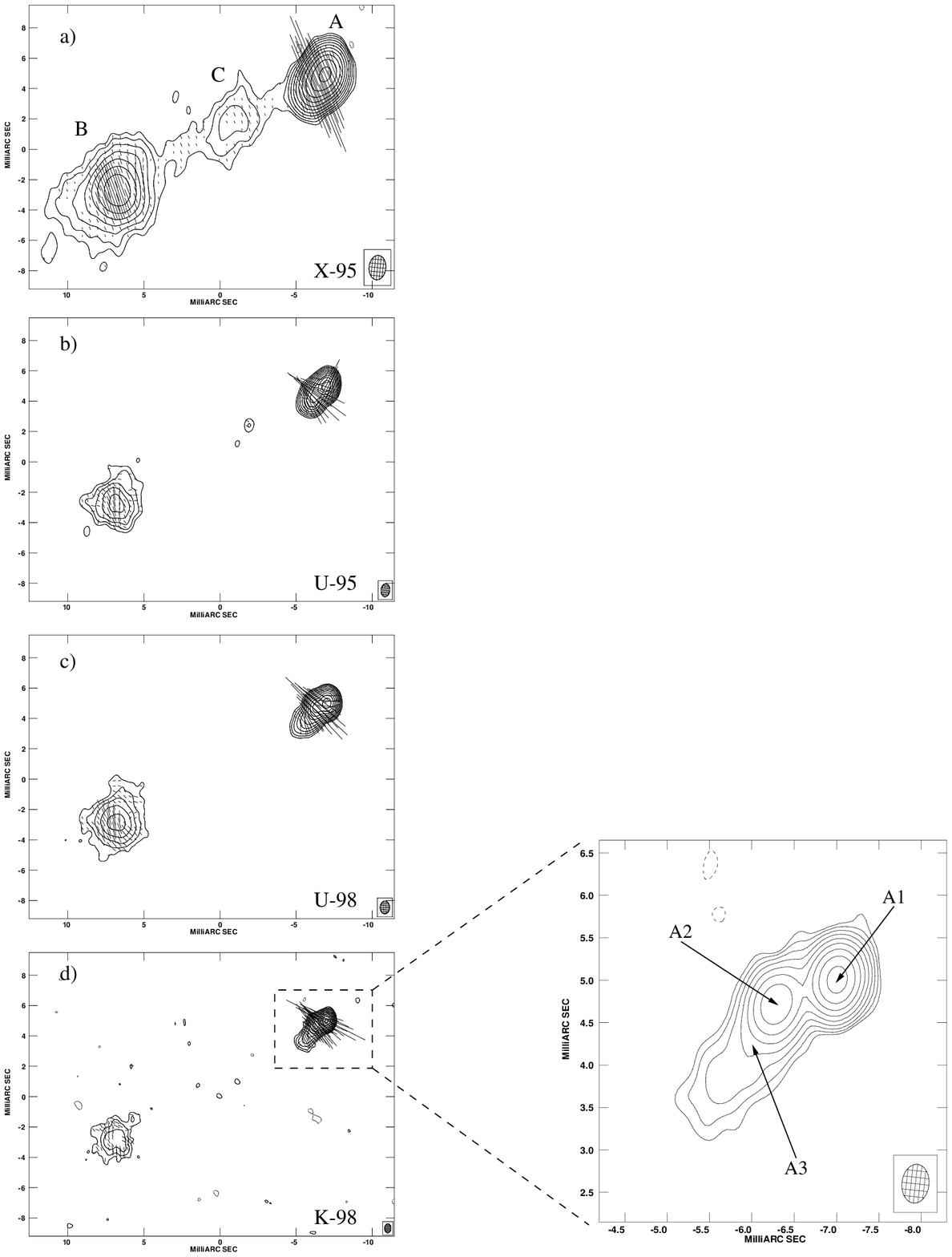}
%\rule{0.4pt}{4cm}% line thickness, height of picture
\vspace{-23cm}
\hfill     
\parbox[t]{8cm}{\caption[]{
%\caption{
{\bf a - d}, from top to bottom, VLBA maps of 3C\,395 made at 8.4 GHz 
and 15.4 GHz in 1995 (X-95 and U-95, respectively), and at 15.4 GHz and 22.2 GHz in 1998 (U-98 and K-98, respectively). 
In all maps contours are spaced by  factors of 2 in brightness.  
The vectors represent the polarization  position angle (E-vector),  
with length  proportional to the polarized flux. The maps have 
been obtained applying natural weighting to the uv data. 
Fig.~1d shows an enlargement of the innermost jet, obtained after 
applying
uniform weighting to the 22.2 GHz data, with components labeled according to
Section 3.1. For each map we list 
the Gaussian beam size (in mas), the first contour level 
(mJy\,beam$^{-1}$), the peak  of brightness (Jy\,beam$^{-1}$) and the
polarized flux corresponding to 1 mas E-vector length (mJy\,beam$^{-1}$).\\
{\bf a:} Beam = 1.6$\times$1.1 P.A. -5.3$^{\circ}$; 1st cntr = 0.80; 
Peak = 1.017; 1 mas $\equiv$ 4.0 \\
{\bf b:} Beam = 0.85$\times$0.58 P.A. -6.6$^{\circ}$; 1st cntr = 0.80; 
Peak = 0.683; 1 mas $\equiv$ 3.33\\
{\bf c:} Beam = 0.84$\times$0.62 P.A. -1.8$^{\circ}$; 1st cntr = 0.80; 
Peak = 0.612; 1 mas $\equiv$ 2.5\\
{\bf d:} Beam = 0.59$\times$0.42 P.A. -3.9$^{\circ}$; 1st cntr = 0.94; 
Peak = 0.548; 1 mas $\equiv$ 2.5\\
{\bf d (zoom):} Beam = 0.46$\times$0.32 P.A. -6.4$^{\circ}$; 1st cntr = 1.50; 
Peak = 0.53
%}
\label{fig1}
}}
\vspace{15cm}
\end{figure*}

\section{Results}

Radio images of the compact structure of 3C\,395 are shown in
Fig.~\ref{fig1}. They have been obtained applying natural weighting
to the data. Total intensity maps are displayed with superimposed
vectors representing the electric field.

\subsection{Total intensity}

At 8.4 GHz, we can identify the ``classical'' components A, B, and C
(Fig.~\ref{fig1}a), although A shows an elongated structure in S-E
direction as an indication of its now known complex composition. The
jet can be traced continuously from A to B, although we find no
evidence of the radio jet beyond the latter component (Saikia
et al. \cite{saikia}; Lara et al. \cite{lara2}, \cite{lara3}). There
is a sharp decrease in the intensity profile beyond component A. The
total flux density recovered in the VLBI map is $1.70\pm 0.09$ Jy.

The images at 15.4 GHz from 1995 and 1998 (Fig.~\ref{fig1}b-c) show
only emission from components A and B, while component C is
resolved out and is too faint to be detected.  The observations at
this frequency, separated by almost three years, confirm the constant
separation between components A and B.  Moreover, the structure of B
has remained essentially invariable during this period of time.  On
the other hand, A shows a bent core-jet structure with clear changes
between the two epochs, probably resulting from a new ejected
component (see also Fig.~\ref{fig3}). The total flux density
recovered in the VLBI maps at 15.4 GHz is the same in both epochs,
$1.12\pm 0.06$ Jy, despite the structural variations in component A. 
Close-in-time space-VLBI observations at 4.85 GHz
(1 May 1998; Lara et al. \cite{lara3}), which provide an angular
resolution comparable to our 15.4 GHz observations, show a similar
bent core-jet structure.

At 22.2 GHz (Fig.~\ref{fig1}d), the detailed structure of component A
is even more evident, consisting of an unresolved component at the
western end of the brightness distribution (assumed to be the true
core), a new component at a distance of 0.8 mas from the core at
position angle P.A. $\sim 110^{\circ}$ and a jet-like feature directed
along P.A.$\sim 130^{\circ}-140^{\circ}$. The flux density in the VLBI
map is $0.94\pm 0.05$ Jy.

We have fitted simple elliptical Gaussian components to the visibility
data using a least square algorithm within the Difmap package in order 
to obtain a
quantitative description of the milliarcsecond structure of
3C\,395. Parameters describing the Gaussian components for each epoch
and frequency are given in Table~\ref{tab1}. Component A requires a
rather complex model, three Gaussian components, to satisfactorily
reproduce the uv-data: A1 stands for the core; A2 describes the new
component which is clearly separated from the core in 1998.50; A3
describes the outer jet emission beyond component A2 (see
Fig.~\ref{fig1}d). Component B is described in terms of 2 Gaussian
components to account for its compact and extended emission,
respectively. Finally, component C, only detected at 8.4 GHz, is
rather weak and extended and it is represented in terms of a single
and elongated Gaussian component. We also include in Table~\ref{tab1}
a similar fit obtained at 4.85 GHz in 1998.33 (Lara et
al. \cite{lara3}), but modified to describe B with two components,
instead of the three used in the quoted publication.

\begin{table}[t]
\caption[]{Gaussian components derived from the model fitting}
\label{tab1}
\begin{scriptsize}
\begin{tabular}{ccrrrrrr}
\hline
     & ID$^a$ & S$^b$ & D$^c$ & P.A.$^d$ & L$^e$ & r$^f$ & $\Phi^g$ \\
     &        &(mJy)  & (mas) & ($^{\circ}$)~~& (mas) &   & ($^{\circ}$) \\
\hline       						    
A1   & X95 & 741 &  --   & --  &  0.22 &    ? & 133 \\
     & U95 & 691 &  --   & --  &  0.20 & 0.17 & 106 \\
     & U98 & 617 &  --   & --  &  0.16 & 0.28 & 115 \\
     & K98 & 570 &  --   & --  &  0.14 &    ? & 107 \\
     & C98 & 124 &  --   & --  &  0.59 &    ? & 106 \\
\\          	      	      			 	    
A2   & X95 & 353 &  0.56 & 113 &  0.63 & 0.12 & 134 \\ 
     & U95 & 179 &  0.61 & 115 &  0.56 & 0.30 & 132 \\
     & U98 & 314 &  0.76 & 112 &  0.32 & 0.49 & 131 \\
     & K98 & 211 &  0.77 & 111 &  0.26 & 0.49 & 129 \\
     & C98 & 629 &  0.66 & 109 &  0.30 & 0.74 & 132 \\
\\          	      	      				    
A3   & X95 & 323 &  1.25 & 135 &  0.66 & 0.50 & 119 \\
     & U95 & 142 &  1.30 & 135 &  0.50 & 0.59 & 101 \\
     & U98 &  73 &  1.50 & 132 &  1.09 & 0.38 & 131 \\
     & K98 &  84 &  1.06 & 126 &  1.71 & 0.23 & 140 \\
     & C98 & 245 &  1.57 & 126 &  0.58 & 0.71 & 160 \\
\\          	      	      				    
C    & X95 &  25 &  6.97 & 119 &  4.60 & 0.44 & 127 \\
     & C98 &  48 &  7.60 & 118 &  8.70 & 0.22 & 118 \\
\\          	      	      				    
B1   & X95 & 168 & 15.86 & 120 &  1.46 & 0.80 &  38 \\
     & U95 &  71 & 15.88 & 120 &  1.48 & 0.77 &  45 \\
     & U98 &  72 & 15.92 & 120 &  1.53 & 0.79 &  64 \\
     & K98 &  31 & 15.96 & 120 &  1.62 & 0.57 &  53 \\
     & C98 & 270 & 16.00 & 120 &  1.54 & 0.97 &  25 \\ 
\\          	      	      				    
B2   & X95 &  82 & 16.15 & 118 &  4.47 & 0.55 & 138 \\
     & U95 &  30 & 15.81 & 117 &  3.55 & 0.40 & 137 \\
     & U98 &  28 & 15.81 & 118 &  3.95 & 0.46 & 150 \\
     & K98 &  34 & 15.78 & 119 &  2.63 & 0.47 & 137 \\
     & C98 &  79 & 16.83 & 117 &  6.18 & 0.40 & 137 \\
\hline
\end{tabular}
\parbox{8.5cm}{\scriptsize{
$^a$ Frequency and year of observation. C, X, U and K stand for 4.85, 8.4, 15.4 and 22.2 GHz, respectively.\\
$^b$ Flux density of the Gaussian component.\\
$^c$ Angular distance from the western-most component A1.\\
$^d$ Position angle with respect to A1, measured north through east.\\
$^e$ Length of the major axis of the Gaussian component.\\
$^f$ Ratio between the major and minor axis of the Gaussian component.\\
$^g$ Orientation of the major axis, defined in the same sense as the position angle.\\
A question-mark indicates that the parameter involved cannot be well constrained by our data. 
}}
\end{scriptsize}
\end{table}

\begin{figure}
\vspace{7cm}
\includegraphics{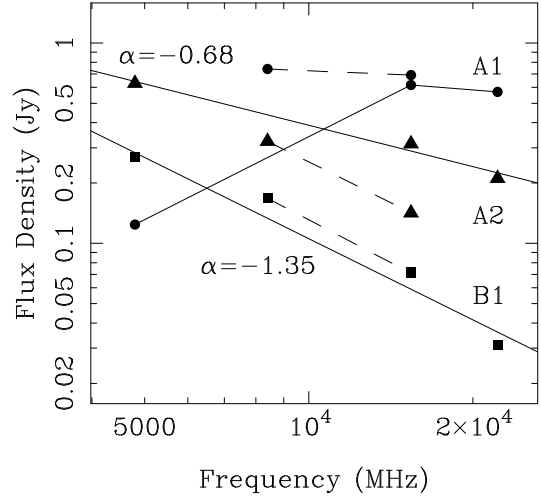}
%%\rule{0.4pt}{4cm}% line thickness, height of picture
\caption{Spectrum of components A1 (circles), A2 (triangles) and B1 
(squares) of 3C\,395 at epoch 1995 (dashed line) and epoch 1998 (solid
line).  See Table~\ref{tab1} for numerical values.  The lines
represent linear fits to the data of each component, excepting A1, for
which the line helps to visualize the spectral dependence of its flux
density. Spectral index values of B1 and A2 correspond to epoch 1998.}
\label{fig2}
\end{figure}

If we combine the close-in-time data at epoch 1998 (4.85 GHz on
1998.33 and 15.4/22.2 GHz on 1998.50) with the results shown in Table
1, we can derive the spectra for the different components of 3C\,395
(Fig.~\ref{fig2}). The spectral decomposition shows that the core
peaks at a frequency between 5 and 15 GHz in 1998. On the other hand, the
other jet components (A2, B1) show a typical steep spectrum with
spectral index $\alpha_{A2}=-0.68$, $\alpha_{B1}=-1.35$ (the spectral
index $\alpha$ is defined so that the flux density $S\propto
\nu^{\alpha}$). Considering the data at 8.4/15.4 GHz on 1995.91, we
can confirm the steep spectrum for A2 and B1, and a flat spectrum for
the core ($\alpha_{A1}=-0.12$), consistent with a shift of the 
turnover frequency towards higher frequencies at the later epoch. 
It is interesting to note that, in
both epochs, the spectrum steepens with increasing core separation.

The components A3 and B2 show also steep spectra in both epochs. We
have not included them in Fig.~\ref{fig2} since they fit the extended
emission of the inner jet and have different angular sizes in our
model fitting, depending on the observing frequency.

\subsection{Polarized emission}

The polarized emission of 3C\,395 comes predominantly from component
A. The orientation of the magnetic field is essentially aligned with
the radio jet, except at the compact core (A1), where we observe a
strong dependence of the polarized intensity and of the magnetic field
orientation with the frequency and epoch of observations. We further
discuss this issue in the next section.

The polarized structure of component B is complex. Since this
component is probably the result of the Doppler boosting of the
radiation from a portion of a relativistic jet which is sharply
directed towards the observer (Lara et al. \cite{lara1}), the changes
in the orientation of the magnetic vector in this region might reflect
the change of orientation of the jet trajectory with respect to the
line of sight. It is interesting to note that the configuration of the
magnetic field suggests that the interaction of the jet with the
surrounding medium is not significant, otherwise the magnetic field
would be compressed and disposed perpendicularly to the jet
direction. This result supports the idea that component B is a 
consequence of a geometry effect in the jet. Moreover, the fact that 
component B does not
show any dependence of the magnetic field orientation and magnitude with 
time is consistent with its stationary character.

Component C is only detected at 8.4 GHz with polarized emission
slightly above the noise level. However, we observe that the magnetic
field tends to be parallel to the jet trajectory in this
component. This fact, joined with its large size, does not lend
support to the interpretation of C as a moving shocked component,
favoring the hypothesis of C being mainly result of the underlying jet
emission.

\section{Discussion}

\subsection{Structural variability}

\begin{figure}
\vspace{9cm}
\includegraphics{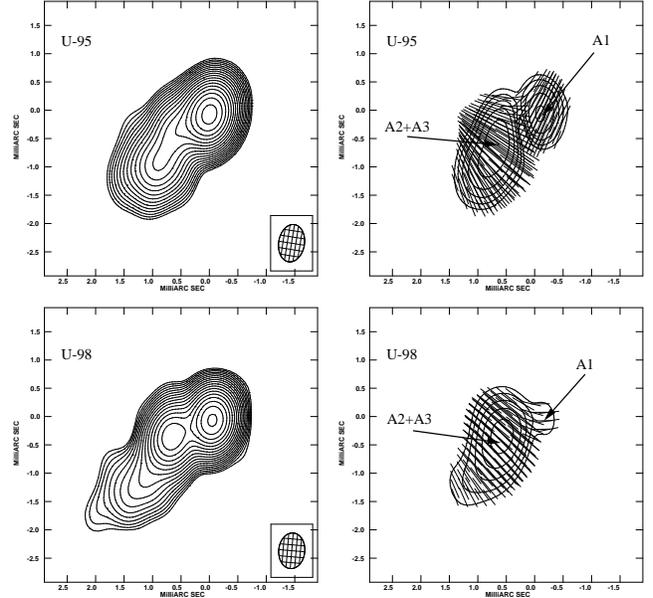}
%%\rule{0.4pt}{4cm}% line thickness, height of picture
\caption{Structural variations in total intensity (left) and polarization 
(right) in the inner structure of 3C\,395, observed at 15.4 GHz
between 1995.91 and 1998.50. The contours are spaced by factors of
$\sqrt{2}$ in all maps. Vectors represent the same as in Fig.~\ref{fig1}. 
For each
map we list the Gaussian beam size (in mas), the first contour level
(mJy\,beam$^{-1}$), the peak of brightness (Jy\,beam$^{-1}$) and in
polarization maps also the polarized flux corresponding to 1
mas E-vector length (mJy\,beam$^{-1}$).  {\bf U-95 left}: Beam =
0.67$\times$0.45 P.A. -9.0$^{\circ}$; 1st cntr = 1.5; Peak =
0.652. {\bf U-95 right}: Beam = 0.94$\times$0.64 P.A. -6.1$^{\circ}$;
1st cntr = 3; Peak = 0.016; 1 mas $\equiv$ 10.  {\bf U-98 left}:
Beam = 0.63$\times$0.46 P.A. -4.9$^{\circ}$; 1st cntr = 1.5; Peak =
0.588; 1 mas $\equiv$ 2.5.  {\bf U-98 right}: Beam = 0.92$\times$0.70
P.A. -1$^{\circ}$; 1st cntr = 3; Peak = 0.014; 1 mas $\equiv$ 10.}
\label{fig3}
\end{figure}

The quasar 3C\,395, considered in the 80's among the fast superluminal
sources (Waak et al. \cite{waak}; Simon et al. \cite{simon}), showed
however a stationary structure in observations made during the 90's
(Lara et al. \cite{lara2}), even though pronounced flux density
variations are usually observed in this source (UMRAO database).  Lara
et al. (\cite{lara3}) found a large bend in the inner region of the
jet, close to the core, and suggested that it could explain why
previous Earth-based cm-VLBI observations did not detect the ejection
of the new moving components expected from the flux density
variability: a large bend in the inner jet would make very
difficult to correlate flux density variability within A with
structural variations beyond this component because the effect of
relativistic time-delay makes the time scales of these two events very
different, and the decrease in the Doppler factor after the bend
produces a large diminution in the flux density of possible moving
components.  As a step forward in the understanding of this source,
the observations presented here show clear evidence of structural
variability close to the core of 3C\,395 between 1995 and 1998, in
agreement with the previous scenario. In Fig.~\ref{fig3} we display
the brightness distribution in total intensity and polarization of 
component A, as observed at 15.4 GHz in 1995.91 and
1998.50, respectively. Changes are evident, both in total intensity
and polarization. The structural variability is consistent with a new
component (labeled A2 in Table~\ref{tab1}) which has been ejected from
the core and is traveling along the jet. From the results of the model
fitting at 15.4 GHz we estimate an apparent velocity of $\beta_{app} =
1.2\pm 0.8 c$ for this component, where the error is calculated
assuming a conservative uncertainty in the position equal to one fifth
of the beam size.

\subsection{Polarization in the inner jet region}

In Table~\ref{tab2} we display, for the different frequencies and
epochs of observations, the total and polarized flux densities, the
mean fractional polarizations and the mean EVPAs of the components of
3C\,395. Total flux densities are taken from Table~\ref{tab1} while
polarized flux densities were measured from the polarization images, 
defining a polygonal area embracing all the component
emission. We were not able to discriminate between the polarized
emission from components A2 and A3, or B1 and B2, since they present
similar EVPA. Thus, we give values for the blending of these
components.

\begin{table}[b]
\caption[]{Polarization in 3C\,395}
\label{tab2}
\begin{scriptsize}
\begin{tabular}{ccrrrr}
\hline
Component & ID$^a$ & S$_{tot}^b$ & S$_{pol}^c$  & $p_m^d$  & $\chi^e$ \\
          &        &  (mJy)    & (mJy)      & (\%)   & ($^{\circ}$)    \\
\hline
A1+A2+A3  & X95 & 1417 & 49 &  3.5 &  22  \\
\\
A1        & U95 &  691 & 11 &  1.6 & 158  \\
A1        & U98 &  617 &  3 &  0.5 & $\sim 93$  \\
A1        & K98 &  570 & 15 &  2.6 &  64  \\
\\
A2+A3     & U95 &  321 & 25 &  7.8 &  40  \\
A2+A3     & U98 &  387 & 24 &  6.2 &  42  \\
A2+A3     & K98 &  295 & 19 &  6.4 &  45  \\
\\
B1+B2     & X95 &  248 & 27 & 10.9 &  19  \\
B1+B2     & U95 &  101 & 10 &  9.9 &  21  \\
B1+B2     & U98 &  100 & 11 & 11.0 &  22  \\
B1+B2     & K98 &   65 &  6 &  9.2 &  15? \\
\hline
\end{tabular}
\parbox{8.5cm}{\scriptsize{
$^a$ Frequency and year of observation: X, U and K stand for 8.4, 15.4 and 22.2 GHz, respectively.\\
$^b$ Total flux density of the component(s).\\
$^c$ Polarized flux density of the component(s).\\
$^d$ Mean fractional polarization.\\
$^e$ Mean EVPA of the component(s).
}}
\end{scriptsize}
\end{table}

The core component, A1, shows strong variability in the polarized
structure at 15.4 GHz (see Fig.~\ref{fig3}). The polarized flux
density of this component decreases from 11 to 3 mJy between 1995.91
and 1998.50.  Although A1 is weak in the latter epoch, we are
confident that it is a real feature in our map, at a level above
$5\sigma$. The EVPA also rotates significantly from 158$^{\circ}$ to
$\sim 93^{\circ}$ in this time period.

\begin{figure}
\vspace{12cm}
\includegraphics{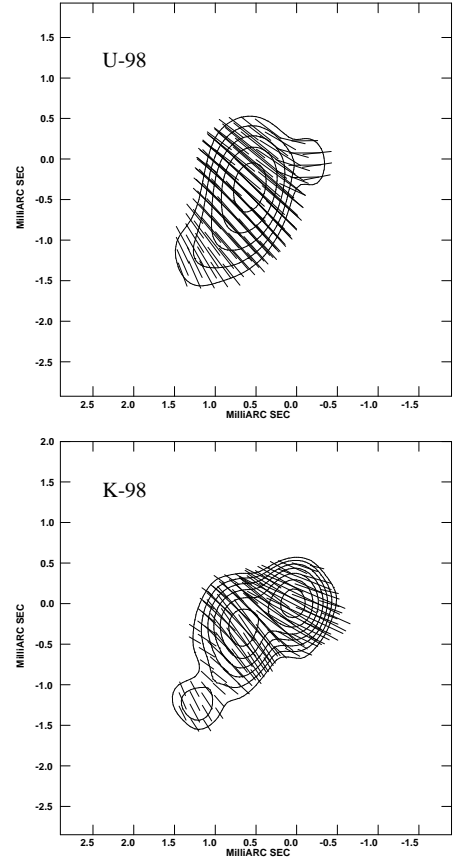}
%%\rule{0.4pt}{4cm}% line thickness, height of picture
\caption{Frequency dependent polarization variations in the very compact 
structure of 3C\,395, observed at 15.4 GHz and 22.2 GHz in 1998.50.
The contours are spaced by factors of
$\sqrt{2}$ in all maps. Vectors represent the same as in Fig.~\ref{fig1}. For each
map we list the Gaussian beam size (in mas), the first contour level
(mJy\,beam$^{-1}$), the peak of brightness (Jy\,beam$^{-1}$) and the 
polarized flux corresponding to 1
mas E-vector length (mJy\,beam$^{-1}$). {\bf Top}: Beam = 0.92$\times$0.70
P.A. -1$^{\circ}$; 1st cntr = 3; Peak = 0.014; 1 mas $\equiv$ 10. {\bf Bottom}: Beam = 0.67$\times$0.51
P.A. -2.9$^{\circ}$; 1st cntr = 1.5; Peak = 0.015; 1 mas $\equiv$ 10.
}
\label{fig4}
\end{figure}

Comparing the results at 15.4 and 22.2 GHz from 1998
(Fig.~\ref{fig4}), we find that component A1 presents a strongly
inverted spectrum in polarization (15 mJy at 22.2 GHz and 3 mJy at
15.4 GHz). Since in 1998 the new component A2 has been completely
ejected from the core region, we suggest that {\em i)} the ``true"
core is almost unpolarized at frequencies lower than 15 GHz, as
expected from synchrotron self-absorption near the core, and {\em ii)}
the polarization variations observed between 1995 and 1998 at 15.4 GHz
in A1 are most plausibly due to the process of ejection of component
A2 and to the resulting opacity changes in the jet. From the 1998 data, 
a RM of $\sim +2500$ rad m$^{-2}$ can be estimated
in component A1, in agreement with Taylor (\cite{taylor1}), which is
consistent with the depolarization of the core region at low
frequencies.  We derive an intrinsic orientation of the electric
vector of $35^{\circ}$, in agreement with the orientation of A2+A3,
and therefore with a magnetic field oriented along the jet.  The
origin of the Faraday rotation is unclear to us. While it might have
an origin internal to the jet, the effects produced by a possible
external ionized screen in the core region cannot be disentangled. We
note that the effect of bandwidth depolarization is negligible with
the frequencies and bandwidths used in our observations, even with the
high RM involved in the core of 3C\,395.

Component A2+A3 shows an EVPA which has remained essentially
constant ($\sim 40^{\circ}$) at 15.4 GHz from 1995 to 1998. On the
other hand, the overall degree of polarization decreases from $\sim
8\%$ to $\sim 6\%$ in this time period. Moreover, the gradual rotation of 
the electric vector along component
A2+A3 is also noticeable (see Fig.~\ref{fig3}). This is probably related 
to the
curvature present in the jet, as suggested by Lara et
al. (\cite{lara3}). At 8.4 GHz it is not possible to discern the
different components within A in polarization. However, since the peak
of the polarized emission is coincident with the position of component
A2 and assuming that A1 is almost unpolarized at this frequency, we
can ascribe the observed EVPA to A2+A3. Comparing with results from 1995 
at 15.4 GHz we estimate for this component a RM of $\sim -350$
rad~m$^{-2}$, which is consistent with the small variation of the EVPA
observed between 15.4 and 22.2 GHz in 1998.

If we interpret the moving component A2 as a planar shock wave, then
the compression of the shock enhances the component of the underlying
magnetic field parallel to the shock front (transverse to the jet
trajectory). This could explain the different field orientation of the core
component A1 with respect to A2+A3 at 15.4 GHz in 1995. However, since
the orientation of the magnetic field in 1998 is along the jet (once
corrected of Faraday rotation), we should conclude that the
enhancement of the field produced by the shock wave is not strong
enough to dominate over the underlying parallel field component. This
is consistent with the slight decrease in the degree of polarization
at 15.4 GHz observed between 1995 and 1998, and in agreement with the
general tendency in quasars to have a net magnetic field orientation
parallel to the jet (Cawthorne et al. \cite{cawthorne}; Wardle
\cite{wardle}).

Finally, we remark the presence of strong gradient in the RM along
the inner jet of 3C\,395, in agreement with Taylor
(\cite{taylor1}). Strong gradients in relativistic jets have also been
found in other compact sources like OJ\,287 (Gabuzda \&
G\'omez \cite{gabuzda}) and BL-Lac (Reynolds \& Cawthorne
\cite{reynolds}).

\subsection{Polarization in component B}

Component B1+B2 has a degree of polarization similar for all
frequencies and observing epochs.  That indicates that Faraday
rotation is not significant in this region of 3C\,395. We note that
Taylor (\cite{taylor1}) finds a moderate RM of $68 \pm 40$ rad m$^{-2}$
in component B, which is fully consistent with our results. Moreover,
the observed EVPA implies that the magnetic field is oriented along
the jet.  As previously mentioned (Section 3.2), the configuration of
the magnetic field suggests that this component is not the result of a
possible interaction of the jet with the external medium, and argues
in favor of a change of the jet geometry.

\section{Conclusions}

We have made VLBA observations of the quasar 3C\,395 at 8.4 and 15.4
in 1995.91, and at 15.4 and 22.2 in 1998.50, and have obtained
detailed maps of its parsec-scale jet, in total intensity and in
polarization. The one-sided jet is complex, showing the previously reported 
gross structure of three components, A, B and C, but with remarkable
sub-structure when observed at sub-milliarcsecond angular resolution.
The emission of 3C395 is strongly dominated by component A.  The new
data confirm the stationary character of component B, also in
polarized emission. Component C appears very elongated at 8.4 GHz,
being more likely the result of the underlying jet emission. It is
resolved out at higher frequencies.

We have studied the sub-structure within component A, i.e. the core
and the inner jet region (of $\sim 3$ mas length), which is crucial to
understand the relation between flux density and structural
variability in 3C\,395.  We have found clear structural variations in
the inner jet, with a new component ejected from the core. In
polarization, we have observed also important time and frequency
dependent variations of the inner jet structure, most possibly 
associated with the ejection of a new component from the core. The 
variability of
this region of 3C\,395 contrasts with the apparent stability
inferred from lower angular resolution observations, which could only
determine that total flux density variability was due to activity
within component A.

The observations show that the jet of 3C\,395 has a projected magnetic
field essentially aligned with the direction of the jet, indicating
that this parallel field is stronger than the field compression
exerted by the shock wave associated with the new ejected component.
The underlying parallel magnetic field is strong enough to be detected
at a distance of 16 mas ($\sim 60$ pc) from the core. We find good
agreement when comparing the degree of polarization and the electric
vector position angle obtained from our VLBA data (Table~\ref{tab2})
and from close-in-time single dish data (database of the University of
Michigan Radio Astronomy Observatory: EVPA 33$^{\circ}$ and
polarization degree of 3\% at 15 GHz in epoch 1998.37), indicating
that the polarization of the source as a whole is mainly associated to
the compact radio source and, in particular, to components A and B.

Finally, we estimate a high Faraday RM, $\sim +2500$ rad~m$^{-2}$,
close to the core region, and a strong gradient along the inner jet,
reaching RM $\sim -350$ rad~m$^{-2}$ at a distance of only 2 - 3 mas
from the core. These values are in agreement with the results derived
by Taylor (\cite{taylor1}).

Follow up observations of moving components like A2 in the inner jet
of 3C\,395 at high angular resolution, by means of high frequency
and/or space-VLBI observations, would eventually allow to trace the
jet curvature close to the core before those components fade away
below the noise level as they reach the sharp cut in the emission
profile beyond component A. We guess that only very strong components
would later be detected in the region between components A and B,
probably like the one observed during the 80's (Waak et
al. \cite{waak}).

\begin{acknowledgements}

We thank Kari Lepp\"anen for his invaluable help during the 1995
data reduction process and the referee for helpful and constructive
comments to the paper. A.A. acknowledges the European Commission's TMR
Programme, Access to Large Scale Facilities, under contract
No. ERBFMGECT950012 for providing financial support to visit the Joint
Institute for VLBI in Europe (JIVE) for data reduction. This research
is supported in part by the Spanish DGES Grant (PB97-1164) and has
made use of data from the University of Michigan Radio Astronomy
Observatory which is supported by funds from the University of
Michigan. The National Radio Astronomy Observatory is a facility of
the National Science Foundation operated under cooperative agreement
by Associated Universities, Inc.

\end{acknowledgements}


\begin{thebibliography}{Normandin \& Kronberg 1980}

\bibitem[1993]{cawthorne} Cawthorne T.V., Wardle J.F.C., Roberts D.H., Gabuzda D.C., 1993, \apj, 416, 519

\bibitem[2000]{gabuzda} Gabuzda D. \& G\'omez J.L., 2000, \mn, in press

\bibitem[1994]{lara1} Lara L., Alberdi A., Marcaide J.M., Muxlow T.W.B., 
1994, \aa, 285, 393

\bibitem[1997]{lara2} Lara L., Muxlow T.W.B., Alberdi A., Marcaide J.M., Junor W., Saikia D.J., 1997, \aa, 319, 405

\bibitem[1999]{lara3} Lara L., Alberdi A., Marcaide J.M., Muxlow T.W.B., 
1999, \aa, 352, 443

\bibitem[1995]{leppanen} Lepp\"anen K., Zensus J.A., Diamond P.J., 1995, \anj, 110, 2479

\bibitem[2000]{reynolds} Reynolds C. \& Cawthorne T.V.  2000, \mn, submitted

\bibitem[1990]{saikia} Saikia D.J., Muxlow T.W.B., Junor W., 1990, \mn, 245, 
503

\bibitem[1994]{shepherd} Shepherd M.C., Pearson T.J., Taylor G.B, 1994, 
\baas, 26, 987

\bibitem[1988]{simon} Simon R.S., Hall J., Johnston K.J., Spencer J.H., Waak J.A., Mutel R.L., 1988, \apj, 326, L5 

\bibitem[2000]{taylor1} Taylor G.B., 2000, \apj, 533, 95

\bibitem[2000]{taylor2} Taylor G.B. \& Myers S.T., 2000, VLBA Scientific Memorandum n.26

\bibitem[1985]{waak} Waak J.A., Spencer J.H., Johnston K.J., Simon R.S., 
1985, \anj, 90(10), 1989

\bibitem[1998]{wardle} Wardle J.F.C. , 1998, in IAU Colloquium 164: Radio
Emission from Galactic and Extragalactic Compact Sources, ASP Conference
Series, J.A. Zensus, G.B. Taylor \& J.M. Wrobel (eds.), Vol. 144, 97

\end{thebibliography}
\end{document}